\documentclass[twoside]{article}
\usepackage{fleqn,MEM_theta}

\usepackage{graphicx}
\usepackage[figuresright]{rotating}

\newcommand{\AmS}{{\protect\the\textfont2
  A\kern-.1667em\lower.5ex\hbox{M}\kern-.125emS}}
 
\newcommand{\simg}{\hspace{.3em}\raisebox{.4ex}{$>$}\hspace{-.75em}
\raisebox{-.7ex}{$\sim$}\hspace{.3em}} 

\title{Application of Maximum Entropy Method to Lattice Field Theory
        with a Topological Term\thanks{Talk presented by
        Y. Shinno}\thanks{~SAGA-HE-201, YAMAGATA-HEP-03-29}}
\author{Masahiro Imachi\address[DPYU]{Department of Physics, Yamagata
        University, Yamagata, Japan}, 
	Yasuhiko Shinno\address[GSSESU]{Graduate School of Science and
        Engineering, Saga University, Saga, Japan}%
	 and 
	Hiroshi Yoneyama\address[DPSU]{Department of Physics, Saga
        University, Saga, Japan}}
       
\begin{document}

\begin{abstract}
 In Monte Carlo simulation, lattice field theory with a $\theta$ term
 suffers from the sign problem. 
 This problem can be circumvented by Fourier-transforming the
 topological charge distribution $P(Q)$. Although
 this strategy works well for small lattice volume, effect of errors of 
 $P(Q)$ becomes serious with increasing volume and prevents one from
 studying the phase structure. This is called flattening. As an
 alternative approach, we apply the maximum entropy method (MEM) to
 the Gaussian $P(Q)$. It is found that the
 flattening could be much improved by use of the MEM. 
\vspace{1pc}
\end{abstract}

\maketitle

\section{INTRODUCTION}
\vspace*{-1mm}
 It is well known that non-perturbative properties of the strong
 interaction are relevant to the dynamics at low energy such as the
 U(1) problem. Although a
 $\theta$ term is deeply associated with the 
 non-perturbative properties, it is indicated experimentally that the
 effect of the $\theta$ term is suppressed. This is the strong CP
 problem. The existence of 
 the $\theta$ term also opens the possibility of rich phase structures
 in $\theta$ space. So it is important to study the dynamics
 of QCD with the $\theta$ term. 
 \par
 The $\theta$ term makes Boltzmann weight complex in the euclidean
 path integral formalism. This makes it difficult to perform Monte Carlo
 simulation. This problem can be circumvented by Fourier-transforming the
 topological charge distribution $P(Q)$. The partition function ${\cal
 Z}(\theta)$ is given as
\begin{equation}
 {\cal Z}(\theta)\equiv\sum_Q P(Q)e^{i\theta Q},
\end{equation}
\vspace*{-1mm}
 where $P(Q)$ is given as 
\begin{equation}
 P(Q)\equiv\frac{\int[{\cal D}\phi]_Q e^{-S(\phi)}}{\int{\cal D}\phi 
  e^{-S(\phi)}}.
\end{equation}
The measure $[{\cal D}\phi]_Q$ represents that the integral is
 restricted to configurations of the field $\phi$ with the topological
 charge $Q$, and $S$ denotes an action.
 \par
 Although this algorithm works well for small lattice
 volume\cite{rf:BRSW,rf:Wiese,rf:HITY,rf:BISY}, 
 the effect of errors of $P(Q)$ becomes serious and disturbs the
 behavior of ${\cal Z}(\theta)$ as volume increases. In fact, a fictitious
 signal of a phase transition was observed due to the errors of
 $P(Q)$\cite{rf:PS,rf:IKY}. This is called flattening. The
 flattening could be 
 remedied by increasing statistics, but it is hopeless because
 exponentially increasing statistics are needed with increasing
 volume. So we do need an alternative way$^{\cite{rf:ADGL}}$ to calculate 
 ${\cal Z}(\theta)$ properly. 
 \par
 In this talk, we use the inverse Fourier transform and apply the
 maximum entropy method (MEM). As a model, we employ the Gaussian
 $P(Q)$.
 \par

\section{MODEL AND FLATTENING}
 The Gaussian $P(Q)$ is parametrized as follows;
\begin{equation}
 P(Q)\propto e^{-\frac{c}{V}Q^2}.\;\;\;(c,V:\mbox{parameters})
\end{equation}
 This model realizes in the 2-d U(1)
 gauge model, the strong coupling limit of the CP$^{N-1}$ model and so
 on. The parameters $c$ and $V$ are regarded as a constant depending on
 the coupling constant and a volume in the U(1) gauge model,
 respectively. For our analysis, we use mock data by adding the Gaussian
 noise with the variance of $\delta\times P(Q)$ to
 the Gaussian $P(Q)$. 
 \par
\begin{figure}[htb]
\vspace*{-4mm}
\centerline{\includegraphics[width=6cm,height=55mm]{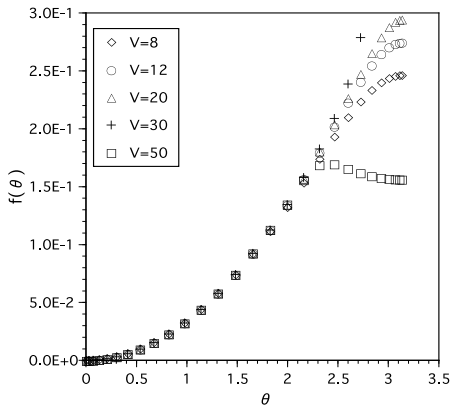}}
\vspace*{-7mm}
\caption{Free energy density $f(\theta)$ for various
 volumes, $c=7.42$ and $\delta=1/400$.}
\vspace*{-5mm}
\label{fig:free}
\end{figure}

 Fig.~\ref{fig:free} shows the free energy density,
 $f(\theta)\equiv-\frac{1}{V}\log{\cal Z}(\theta)$, obtained by
 Fourier-transforming numerically the mock data for various volumes.
 \par
 As $V$ increases, already at $V=30$, $f(\theta)$ cannot be calculated
 correctly because of errors of $P(Q)$ in Fig.~\ref{fig:free}.
 Especially at $V=50$, $f(\theta)$ becomes flat for $\theta\simg
 2.3$ and gives a fictitious signal of a first-order phase transition at
 $\theta\simeq 2.3$. This is nothing but the flattening. Since the
 flattening is characteristic of the Fourier transform procedure, an
 alternative way should be employed to calculate $f(\theta)$ properly. 
 \par

\section{MAXIMUM ENTROPY METHOD}
 In this talk, the inverse Fourier transform is used in the
 analysis. In this case, the number of degrees of freedom of $Q$, $N_q$,
 is smaller than that in $\theta$ space, $N_\theta$, and the analysis by
 use of the $\chi^2$-fit suffers from the ill-posed problem. In order
 to circumvent this problem, we use the MEM,
 which is effective for such the issue. 
 \par
 The MEM is based on the Bayes' theorem in the probability theory. For
 our case, 
 probability ${\rm prob}({\cal Z}(\theta)|P(Q),I)$ is considered,
 which is the probability that ${\cal Z}(\theta)$ is realized when data
 of $\{P(Q)\}$ 
 and information $I$ are given. The information $I$ represents our state of
 knowledge about ${\cal Z}(\theta)$. In this case, we impose a 
 criterion that ${\cal Z}(\theta)>0$. 
 \par
 The probability ${\rm prob}({\cal Z}(\theta)|P(Q),I)$ is written in terms of
 $\chi^2$ and the entropy $S$; 
\begin{equation}
 {\rm prob}({\cal Z}(\theta)|P(Q),I)\propto \exp\bigl\{-\frac{1}{2}\chi^2+
  \alpha S\bigr\}\equiv e^{W({\cal Z})},
\end{equation}
 where $\alpha$ is a real-positive parameter. Conventionally the
 Shannon-Jaynes entropy $S$ is employed\cite{rf:Bryan,rf:AHN}. 
\begin{equation}
 S=\int^\pi_{-\pi}d\theta\biggl[{\cal Z}(\theta)-m(\theta)
  -{\cal Z}(\theta)\log\frac{{\cal Z}(\theta)}{m(\theta)}\biggr],
\end{equation}
 where $m(\theta)$ is called default model which reflects our state of
 knowledge about ${\cal Z}(\theta)$. Namely, our task is to explore the
 image ${\hat{\cal Z}}(\theta)$ such that ${\rm prob}({\cal
 Z}(\theta)|P(Q),I)$ is maximized by following the three
 steps\cite{rf:Bryan,rf:AHN};  
 \par
\begin{enumerate}
 \item Maximizing $W({\cal Z})$ for a given $\alpha$:
\begin{equation}
 \frac{\delta W({\cal Z})}{\delta {\cal Z}(\theta)}
  \Biggm|_{{\cal Z}={\cal Z}^{(\alpha)}}=0.
\end{equation}
%
 \item Averaging ${\cal Z}^{(\alpha)}(\theta)$ to calculate the best image
       ${\hat{\cal Z}}(\theta)$ which maximizes 
       ${\rm prob}({\cal Z}(\theta)|P(Q),I)$:  
\begin{equation}
 {\hat{\cal Z}}(\theta)\simeq\int d\alpha~{\rm prob}(\alpha|P(Q),I)
  {\cal Z}^{(\alpha)}(\theta),
\end{equation}
 \item Error estimation.
\end{enumerate}
 \par

\section{RESULTS}
 In our analysis, the Gaussian $P(Q)$ is used as a test. It is
 a good laboratory to investigate how the flattening is improved, because
 the corresponding partition function ${\cal Z}_{{\rm pois}}(\theta)$ is
 calculated by use of the Poisson's sum formula analytically. We
 perform the analysis for various 
 parameters $c$, $V$, $\delta$. Here we fix $c=7.42$, $V=50$ and
 $\delta=1/400$. The number of data set is 30. Three default models
 are employed: (i) constant type; $m(\theta)=1.0$, (ii) strong coupling
 region type; $m(\theta)=(\frac{2}{\theta}\sin\frac{\theta}{2})^V$,
 (iii) Gaussian type; $m(\theta)=\exp\bigl\{-\frac{\log 10}{\pi^2}\gamma
 \theta^2\bigr\}$, where the parameter $\gamma$ is varied from 4 to 8.
 In the analysis, it is non-trivial to find a solution due to
 $N_q<N_\theta$ and the singular value decomposition is 
 employed. In order to calculate image
 ${\hat{\cal Z}}(\theta)$ with high precision, the Newton method
 is used with quadruple precision. 
 \par
 Firstly, in order to find a solution which approximately agrees with
 the exact ${\cal Z}_{{\rm pois}}(\theta)$, analyses are performed by
 using the three default models for various $\alpha$ by step 1.  
 \par
\begin{figure}[htb]
\vspace*{-4mm}
\centerline{\includegraphics[width=6cm,height=58mm]{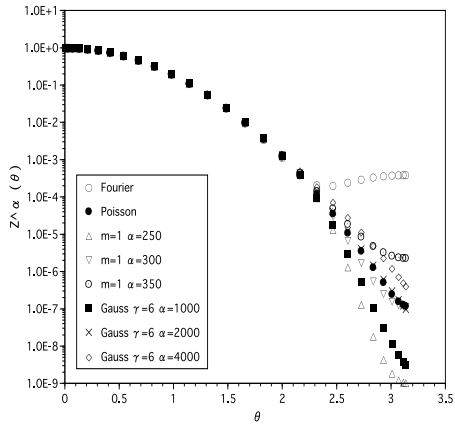}}
\vspace*{-7mm}
\caption{${\cal Z}^{(\alpha)}(\theta)$ for data at $V=50$ with
flattening. As a comparison, the result of Fourier transform($\circ$)
 and ${\cal Z}_{{\rm pois}}(\theta)$($\bullet$) are also displayed.}
\vspace*{-5mm} 
\label{fig:Zalpha}
\end{figure}
 Fig.~\ref{fig:Zalpha} shows the results. Note that all the results of
 the MEM are free from the flattening. Results of the constant type at
 $\alpha=300$ and the Gaussian one with $\gamma=6$ at $\alpha=2000$
 approximately agree with the exact ${\cal Z}_{{\rm pois}}(\theta)$. 
 \par
 In order to investigate whether these solutions are favored
 probabilistically, we calculate ${\hat{\cal Z}}(\theta)$ following 
 step 2 and also estimate these errors by step 3. We find that
 ${\hat{\cal Z}}(\theta)$ given by the Gaussian default model with
 $\gamma=6$ is the most favorite image in the current analysis.
 \par 
\begin{figure}[htb]
\vspace*{-4.2mm}
\centerline{\includegraphics[width=6cm,height=58mm]{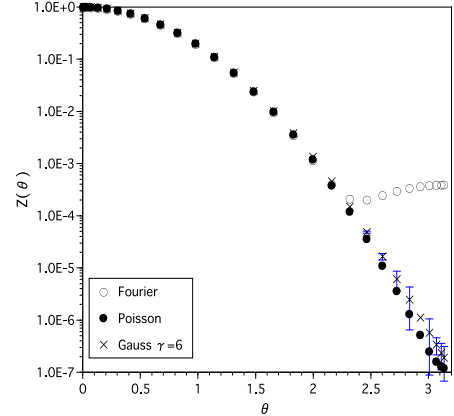}}
\vspace*{-7mm}
\caption{${\hat{\cal Z}}(\theta)$ given by the Gaussian type with
 $\gamma=6$.} 
\vspace*{-5mm} 
\label{fig:Zimage}
\end{figure}
 The result is shown in Fig.~\ref{fig:Zimage}. The result is free from
 the flattening and consistent with 
 ${\cal Z}_{{\rm pois}}(\theta)$. Hence we conclude that the MEM works
 for the $\theta$ term and reproduces the reasonable image\cite{rf:ISY}. 
 \par

\section{SUMMARY}
 In this talk, we applied the MEM to mock data of the Gaussian $P(Q)$ and
 showed that the flattening is much improved by use of the MEM. 
 \par
 The next task is to apply the MEM to more realistic models. 
 \par

\end{document}